\documentclass[useAMS,usenatbib]{mn2e}
\usepackage{epsfig}
\usepackage{times}
\usepackage[fleqn]{amsmath}
\usepackage{amssymb}
\usepackage{amsbsy}

\usepackage[fleqn]{amsmath}
\usepackage{amsfonts}
\usepackage{gensymb}
\usepackage{epstopdf}
\usepackage[bottom]{footmisc}

\newcommand{\be}{\begin{equation}}
\newcommand{\ee}{\end{equation}}
\newcommand{\ba}{\begin{eqnarray}}
\newcommand{\ea}{\end{eqnarray}}

\newcommand{\lya}{\mbox{Ly$\alpha$ }}

\title[Mapping cosmic web filaments in Ly$\alpha$ emission]{Mapping the low surface brightness Universe in the UV band with Ly$\alpha$ emission from IGM filaments}

\author[M. B. Silva et al.]{Marta B. Silva$^1$ \thanks{E-mail:
silva@astro.rug.nl}, Robin Kooistra$^1$, Saleem Zaroubi$^1$ $^2$\\
Kapteyn Astronomical Institute, University of Groningen, Landleven 12, 9747AD Groningen, the Netherlands\\
Department of Natural Sciences, The Open University of Israel, 1 University Road, P.O. Box 808, Ra'anana 4353701, Israel
}

\begin{document}

\maketitle
\begin{abstract}

A large fraction of the baryonic matter in the Universe is located in filaments in the intergalactic medium. However, the low surface 
brightness of these filaments has not yet allowed their direct detection except in very special regions in the circum-galactic 
medium (CGM).

Here we simulate the intensity and spatial fluctuations in Lyman Alpha ${\rm (Ly\alpha)}$ emission from filaments 
in the intergalactic medium (IGM) and discuss the prospects 
for the next generation of space based instruments to detect the low surface brightness universe at UV wavelengths.
Starting with a high resolution N-body simulation we obtain the dark matter density fluctuations and associate baryons 
with the dark matter particles assuming that they follow the same spatial distribution. 
The IGM thermal and ionization state is set by a model of the UV background and by the 
relevant cooling processes for a hydrogen and helium gas.
The \lya emissivity is then estimated, taking into account recombination and collisional excitation processes. We 
find that the detection of these filaments through their \lya emission is well in the 
reach of the next generation of UV space based instruments and so it should be achieved in the next decade. 
The density field is populated with halos and galaxies and their \lya emission is estimated. Galaxies are treated 
as foregrounds and so we discuss methods to reduce their contamination from observational maps. 
Finally, we estimate the UV continuum background as a function of the redshift of the \lya emission line and 
discuss how this continuum can affect observations. 

\end{abstract}


\section{Introduction}
 
Galaxies and galaxy clusters are connected by filamentary structures known as the cosmic web. A large fraction of the 
baryonic matter is located in these low density filaments which are characterized by a very low surface brightness.
Observing and mapping these baryons is an important step in order to complete our understanding of the mechanisms 
involved in galaxy formation and evolution at large scales as well as to correctly model interactions and clustering of galaxies.
Also, the cold gas in these filaments will eventually flow into galaxies and be involved in processes of stellar 
formation and therefore its important to accurately model its thermal state.

While the CGM can be probed with metal lines such as OVI and CIV 
\citep{2004ApJ...606..221F,2014MNRAS.438.1820K,2015arXiv150707002L}, 
lower density regions of the IGM have a much smaller metallicity and can only be probed through hydrogen or 
helium transition lines.
At high redshifts neutral gas in the IGM is often detected through observations of \lya absorption 
features in the spectra of quasars \citep{1965ApJ...142.1633G} or through observations of \lya blobs, which are 
spatially extended Lyman alpha nebulae in the densest regions of the Universe. At lower redshifts the gas, however 
is more ionized and so the scattering of \lya photons is reduced. Searches for \lya blobs at $z\sim 0.8$ with GALEX 
(The Galaxy Evolution Explorer) have so far not been successful, which might indicate that these structures are 
restricted to the high redshift universe \citep{2009AJ....138..986K}.

Also, \lya emission originating at a redshift of $z\lesssim 2.2$ 
has to be detected from space, because the Earth's atmosphere is opaque to UV radiation. At low redshifts the 
neutral gas is therefore mostly probed by radio observations of HI ${\rm 21\, cm}$ absorption features. Current 
radiotelescopes can only probe this line at $z\lesssim0.6$, although in the far future the SKA-2 experiment 
will be able to go much higher in redshift.    

The hydrogen \lya line is usually the strongest emission line in galaxy spectra with an intrinsic flux proportional to 
the ionizing photons emissivity. This is also a promising line to probe the IGM, where it is usually the main cooling mechanism.
 
Lyman alpha emission is usually associated with ionized recombining gas and in the cold IGM its intensity is mainly set 
by the thermal and ionization state of the gas which are related to the UV and X-ray backgrounds. 
The maximum \lya intensity in regions powered by the UV/X-ray backgrounds is about 50$\%$ of the intensity of this ionizing radiation
and can therefore be used to constrain it.
In the overdense CGM, most of the gas, however is shock heated and collisional excitation is the main mechanism leading 
to \lya emission \citep{2003ApJ...599L...1F,2005ApJ...622....7F,2010MNRAS.408.2051D}.
The fraction of the gas that is shock-heated increases towards lower redshift but should be smaller than the cold gas 
fraction, even in the local universe \cite{2010MNRAS.408.2051D}.   
Currently, gas filaments in the IGM are mostly indirectly detected through optical observations of the stars 
and galaxies they contain.
A direct detection of a large gas filament surrounding a quasar was recently reported by \cite{2014Natur.506...63C} at 
$z\sim2.3$. The \lya emission in this filament is powered by UV emission from a local quasar. Thus, its 
intensity is much larger than the expected UV background powered emission. Also, it is likely that this 
filament is at a density peak which is further boosting its \lya emission. 

Here, we propose to use intensity mapping of the \lya line from the local universe to z$\sim$3
as a probe of the IGM baryonic content and of the astrophysical properties of IGM filaments. The 
main objective is to detect and characterise the faint emission from UV/X-ray background powered 
filaments. Additionally, with these maps we aim to probe the UV background intensity, since the 
maximum \lya emissivity from IGM filaments occurs in optically thick regions and should 
therefore have roughly half of the intensity of this background. 

Intensity maps will mainly provide statistical quantities such as the intensity and 
power spectrum of the signal, with the advantage that for this it is not required for the signal to be 
above the instrumental noise, such as in the case of galaxy surveys. 

The possibility of mapping \lya emission in the optical band at $z\, =\, 3$ has already been discussed 
in \cite{2005ApJ...622....7F} and \cite{2014ApJ...786..106M}. In the UV band, most studies of this 
transition are focused on the prospects for mapping Lyman alpha absorbers identified by the mean flux decrement
along galaxy lines of sight (see eg. \cite{2010MNRAS.408.2051D}) and do not consider the possibility of detecting  
\lya emission in the IGM, since it was not seen as an attainable possible goal in the foreseeable future. 
The problem resides mainly in the technical difficulties in measuring UV radiation, while successfully 
removing the galactic and extragalactic contamination.
Fortunately, new technological advances make \lya intensity mapping in the UV an achievable goal in the next 
decade. Therefore, in this study, we use updated astrophysical and cosmological parameters to 
estimate the \lya signal from the IGM. In particular, estimates are shown for the required sensitivity, field-of-view 
and resolution to 
measure the bulk of the signal and to detect the power spectra of \lya emission fluctuations.

Our study shows that any high sensitivity galaxy survey in the UV band will also 
be sensitive to emission from gas filaments. Even if the overall \lya emission from galaxies is 
much higher than the emission from filaments, the effective escape fraction of \lya photons from 
galaxies at $z\simeq 3.0$ is quite small and so it will push towards increasing the new UV instruments, sensitity. 
Conversely, the \lya escape fraction of filaments is close to unity, which makes them easier to observe.

This paper is organized as follows. The model and the simulations used to obtain 
maps of the ionization and thermal state of the IGM are presented in Section \ref{sec:IGM_sim}. In Section \ref{sec:Lya_emission} 
the processes leading to \lya 
emission from galaxies and from the IGM are discussed and predictions for the overall intensities are presented.
In Section \ref{sec:results} the main results are shown, including several maps obtained from the simulation and
the several relevant \lya emission power spectra.
The experimental setup necessary for measuring the target emission is presented in Section \ref{sec:Experiments}. In 
Section \ref{sec:Foregrounds} the modeling of the contaminants in \lya emission intensity maps is 
introduced and foreground removal techniques are discussed.
Finally, the main results and conclusions are presented in Section \ref{sec:Conclusions}.  

\section{Modeling the thermal and ionization state of the IGM}
\label{sec:IGM_sim}

Direct detection of IGM filaments through observations of emission/absorption by transition lines will make it possible to 
properly map the thermal and ionization state of the IGM gas, which can be used to constrain the 
UV/X-ray background originating in the overall emissivity from quasars and stars through time. 
At the relevant redshifts for this study, this background is dominated by emission from quasars and 
so the proposed observations will mainly put constraints on the overall quasar luminosity density 
and, in particular, on the poorly constrained low energy end of the quasars emission spectra.

We employ the Gadget 2 
 N-body code to obtain a dark matter (DM) only simulation \citep{2001NewA....6...79S,2005MNRAS.364.1105S}
using the best fit cosmological parameters from Planck + WMAP \citep{2014A&A...571A..16P} 
($\Omega_b h^2=0.022032$, $h=0.6704$, 
$Y_P=0.2477$, $n_s=0.9619$ and $\sigma_8=0.8347$) and save the outputs correspondent to redshifts 0 to 3.

The simulated volume is ${\rm 50\, Mpc^3/h^3}$ with a ${\rm 10^7\, M_{\odot}}$ mass resolution.
The cloud in cell (CIC) method is then used to distribute the particles in 3D boxes with $N=800^3$ cells.
In this study we assume that the baryon spatial distribution follows that of the dark matter particles.

The temperature and ionizing state of the gas in each cell of the simulation were modeled, assuming 
ionizing and thermal equilibrium set by the ionizing background radiation, adiabatic cooling, and all the heating,
recombination and collisional processes relevant for a hydrogen and helium gas. The temperature of the gas 
was obtained with 

\be
\frac{dT_{\rm gas}}{dt}=-2H(z)T_{\rm gas} + \frac{2}{3}\frac{{\cal H} - \Lambda}{nk_B},
\ee

where ${\cal H}$ and $\Lambda$ are respectively the heating and cooling functions and n is the baryon number density.

The adiabatic cooling was computed assuming only the average Hubble expansion and neglecting peculiar velocities and we are 
therefore underestimating it. Peculiar velocities dominate the gas expansion at low scales. However, at the large scales relevant 
for this study these velocities are smoothed enough that, for overdensities of 10 - 20 times the average gas density, the peculiar 
velocity gradients are on average a few times lower than the average Hubble expansion. For more 
details see (Kooistra et al. in prep). We did not account for metal 
cooling given the expected low metallicity of IGM filaments and the low temperatures involved.

We used the \cite{2012ApJ...746..125H} UV/X-ray background model, which includes the integrated emissivity 
from star forming galaxies and active galactic nuclei (AGN), as a source of heating and ionizing photons. 
This model assumes a relatively small escape fraction of ionizing photons from star forming galaxies and 
therefore there is a good chance that it is underestimating this emission. However, at $z<3$ the predicted 
background is dominated by AGN emission and so we will not discuss the uncertainties in this part of the model further.

The AGN background radiation predicted by this model is based on the few available 
observational results of AGN number counts and spectral emissivities. Attenuation of this background
by absorption in gas clouds in the IGM is also taken into account, based on observations of
\lya forest features in the spectra of background sources.

There is a large uncertainty in the modelling of the AGN background radiation due to the poorly known number density of AGN as we go 
to higher redshifts and due to the lack of measurements of AGN emissivity at UV wavelengths.
The resulting uncertainty in the UV/X-ray background is therefore of at least one order of magnitude and increases towards lower redshifts. However, 
the impact of this uncertainty in the gas ionization and thermal state, and therefore in the total \lya emissivity 
from IGM filaments, will be of a factor of 2 or 3 at most. 

At z$<$3, and especially as we approach z=0, the 
attenuation length is high enough that there are several sources contributing to the flux arriving at each 
point of the IGM and so it is fair to assume a spatially homogeneous UV background \citep{2009RvMP...81.1405M}. 
For the cold IGM, we can assume that the heating and the ionization state of gas filaments is 
set by the UV/X-ray background. However, in the vicinity of a young galaxy or 
an active galactic nucleus, the gas will be hotter and more ionized and so the equilibrium assumptions will not hold.

Also, shock-heating of gas during gravitational collapse of matter is increasingly important towards lower redshifts.
This code does not account for shock-heating of gas, metal cooling or self-shielding of overdense gas.
The correct modeling of these processes requires very high resolution hydrodynamical simulations 
which would restrict this study to low volumes, which is not our objective since we mainly 
propose to probe the large scale structure distribution of gas and the location of the baryons in the cold IGM. 
For that we need relatively large volumes and do not need to properly resolve the complex processes 
involved in setting the temperature and ionization state of the gas in the CGM.
Nevertheless, even in the local Universe most of the IGM baryons are likely to be located in cold 
gas filaments \cite{2010MNRAS.408.2051D}.  
For more precise hydrodynamic simulations of \lya emission/absorption in the IGM we refer the reader to the study of 
fluorescent \lya emission by \cite{2010ApJ...708.1048K} which mainly targets emission from $z\sim 2-3$. Another useful 
reference to which we compare some of our results is the study by \cite{2010MNRAS.408.2051D}, who simulated 
IGM \lya absorbers in a volume of ${\rm 48\, Mpc^3/h^3}$ from $z\sim 0-2$ with the main objective of predicting 
the number of \lya absorbers detectable using the Hubble’s Cosmic Origins Spectrograph.

\begin{figure}
\begin{center}  
\hspace{-4 mm}
\includegraphics[angle=0,width=0.50\textwidth]{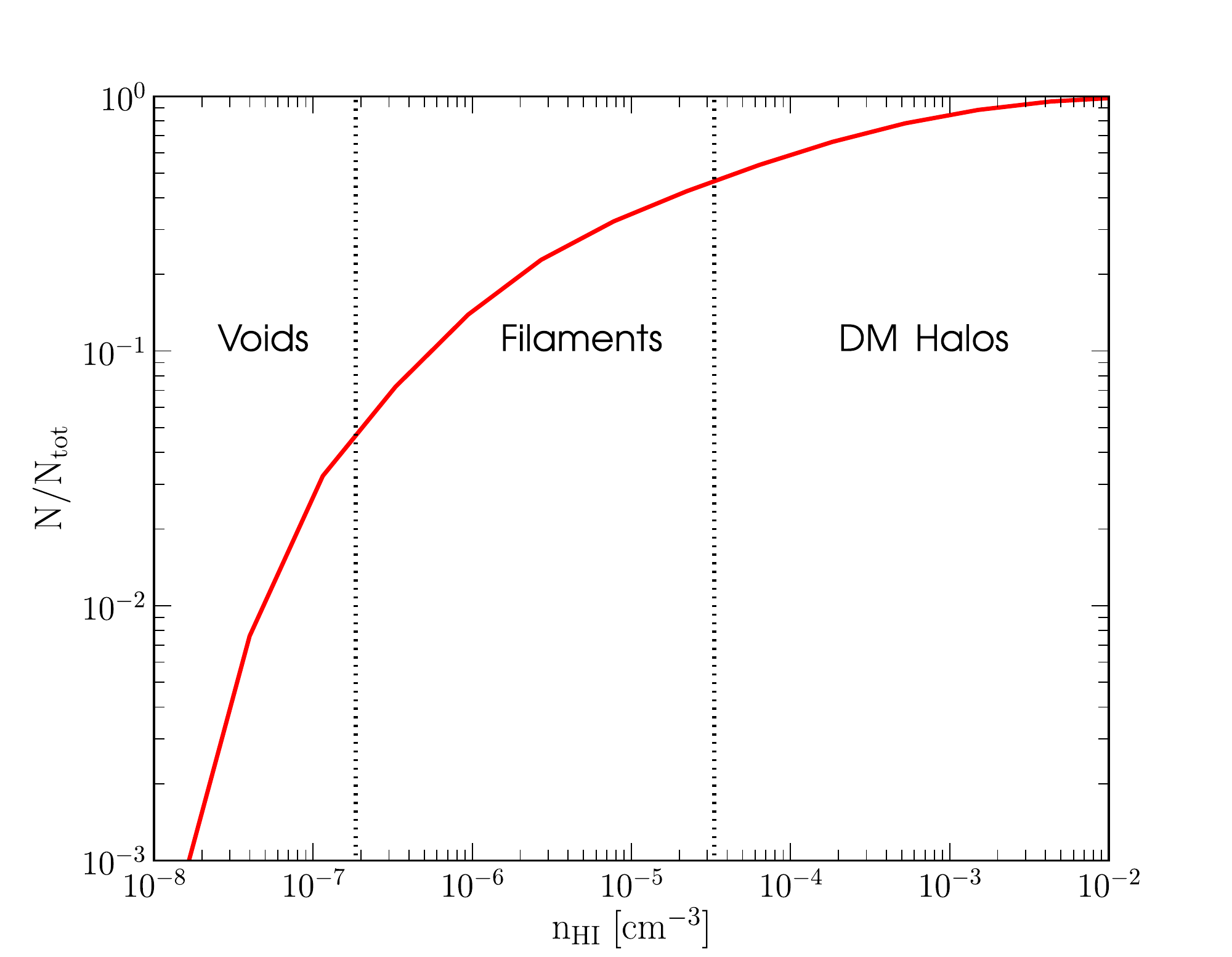}
\caption{Cumulative distribution of hydrogen baryons as a function of matter overdensity at $z=0.65$.}
\label{fig:xHI_z}
\end{center}
\end{figure}

The distribution of hydrogen in galaxies and in the IGM as a function hydrogen density is shown 
in Figure \ref{fig:xHI_z}. This figure shows that a large fraction of the baryonic matter is located 
in IGM filaments and so it illustrates the importance of observing this filaments.

\section{Modeling lyman alpha emission}
\label{sec:Lya_emission}

In this section, we describe the essential assumptions of our modeling of \lya emission 
from the IGM and from galaxies and then
in Section \ref{sec:results} we present our simulation analysis and main results.

Lyman alpha photons are emitted by hydrogen during recombinations or due to 
collisional excitations between neutral hydrogen and free electrons. Both of these processes 
are usually powered by UV/X-ray emission from stars or quasars.
The additional \lya emission, powered by the energy released during gravitational collapse, is mostly 
relevant for studies of the CGM and so it can be safely ignored in this study. On the other hand, an 
additional source of \lya photons would only increase the detectability of the proposed target signal.

\subsection{Lyman alpha emission from the IGM}
\label{sec:model_lya_IGM}

Lyman alpha emission naturally traces the more overdense regions of the gas where the 
recombination and excitation rates for hydrogen are higher, which in the IGM corresponds
to the overdense gas filaments. 
The intensity of \lya emission in these filaments depends on their density, ionization and thermal state.
The luminosity density (per comoving volume) in \lya emission from hydrogen recombinations in the IGM, $\ell^{\rm IGM}_{\rm rec}$ is
\be
{\rm \ell}^{\rm IGM}_{\rm rec}(z)= f_{\rm rec} \dot{n}_{\rm rec}  E_{\rm Ly\alpha},
\label{eq:lya_IGM_rec}
\ee
where the probability of emission of a \lya photon per recombination of a hydrogen atom is
\be
f^{\rm A}_{\rm rec}=0.41-0.165\log_{10}(T_{\rm K}/10^4{\rm K})-0.015(T_{\rm K}/10^4{\rm K})^{-0.44}.
\ee
\begin{figure}
\begin{centering}  
\hspace{-4 pt}
\includegraphics[angle=0,width=0.50\textwidth]{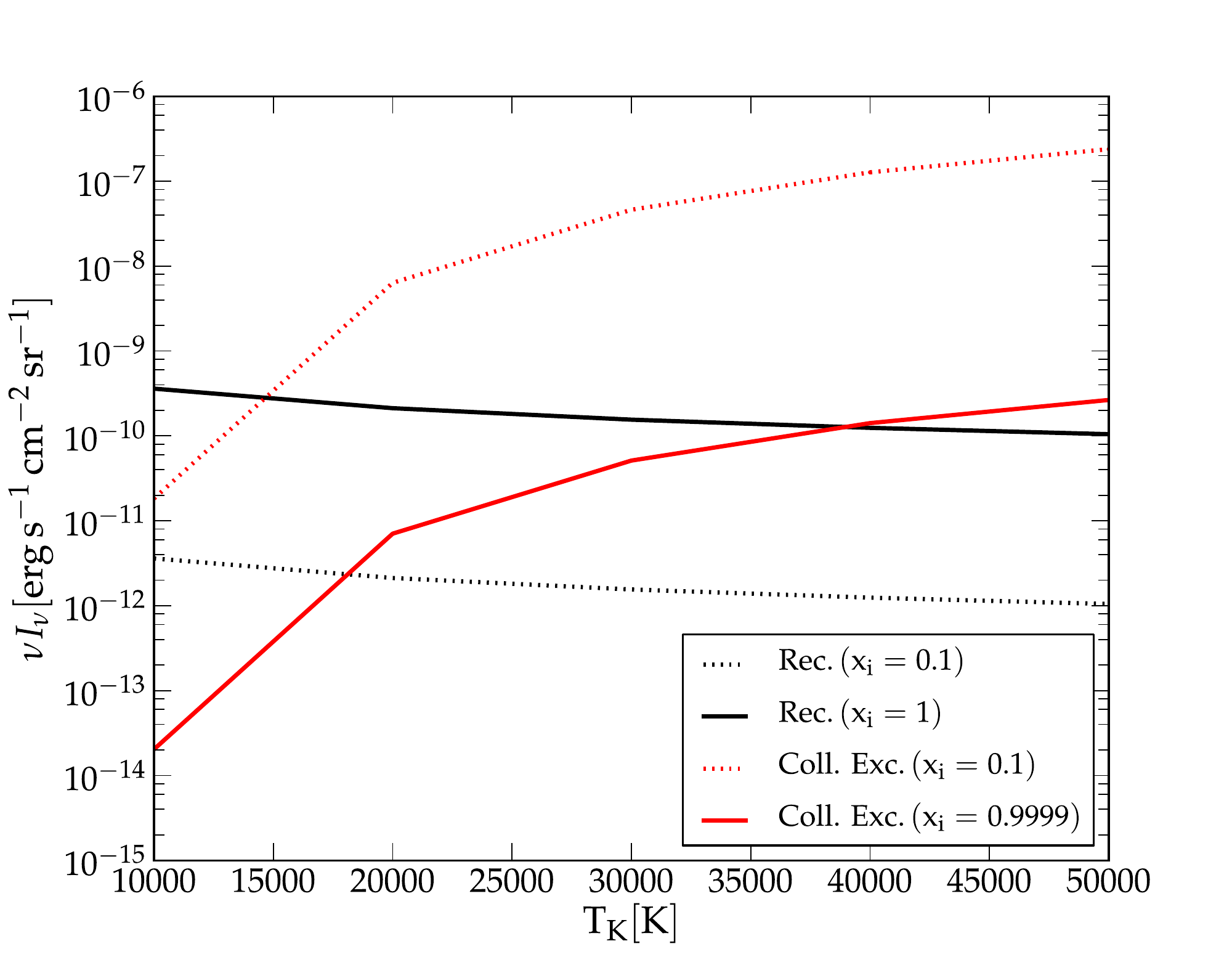}
\caption{Theoretical intensity of \lya emission from recombinations and collisions in the IGM. Recombination 
emission was estimated assuming full ionization of the gas, whereas collisional emission was estimated for different 
values of $x_{\rm i}$. The intensities shown assume a clumping factor of 1 and can be easily scaled to higher clumping 
factors just by multiplying the resulting intensity by C.}
\label{fig:Teo_lya}
\vspace{2pt}
\end{centering}

\end{figure} 
This relation is taken from \cite{2014PASA...31...40D} and is appropriate for $100\, {\rm K}\, <\, T_K\, <\, 10^5\, {\rm K}$ and
for a case A recombination coefficient.
The number density of recombinations per second, $\dot{n}_{rec}$, is given by:
\be
\dot{n}_{rec}(z)=\alpha_A n_e(z) n_{\rm HII}(z),
\label{eq:nrec_s}
\ee
where $n_{\rm HII}=x_i \frac{n_b(1-Y_p)}{1-3/4Y_p}$ is the ionized hydrogen number density 
($x_i$ is the mass averaged ionized fraction, $n_b$ the baryon comoving number density) and 
the free electron density can be approximated by $n_{\rm e}=x_{\rm i} n_{\rm b}$. In the IGM, the gas 
is mostly optically thin and so the recombination probability was modeled with the case A recombination 
coefficient taken from \cite{1994MNRAS.269..563F}

\ba
\alpha_A &\approx& 6.28\times 10^{-11} T_{\rm K} ^{-0.5}(T_{\rm K}/10^3 {\rm K})^{-0.2}\\ \nonumber
&\times&[1+(T_{\rm K}/10^5 {\rm K})^{0.7}]^{-1}\ {\rm cm^3\, s^{-1}}.
\ea
Using a case B recombination rate would result in a smaller number of recombinations but a higher probability 
of \lya photon emission per recombination. Therefore, the overall emission rate of \lya photons has a weak 
dependence on the choice of recombination coefficient \citep{2014PASA...31...40D}.
The luminosity density (per comoving volume) in \lya emission from collisional excitation of neutral 
hydrogen by free electrons in the IGM is
\be
{\rm \ell}^{\rm IGM}_{\rm exc}(z)= n_{\rm e} n_{\rm HI} q_{\rm Ly\alpha} E_{\rm Ly\alpha},
\label{eq:lya_IGM_exc}
\ee
where $q_{\rm Ly\alpha}$ is the effective collisional excitation coefficient which is calculated
as in \cite{2013ApJ...763..132S}.

\begin{figure}
\begin{centering}  
\hspace{-4 pt}
\includegraphics[angle=0,width=0.52\textwidth]{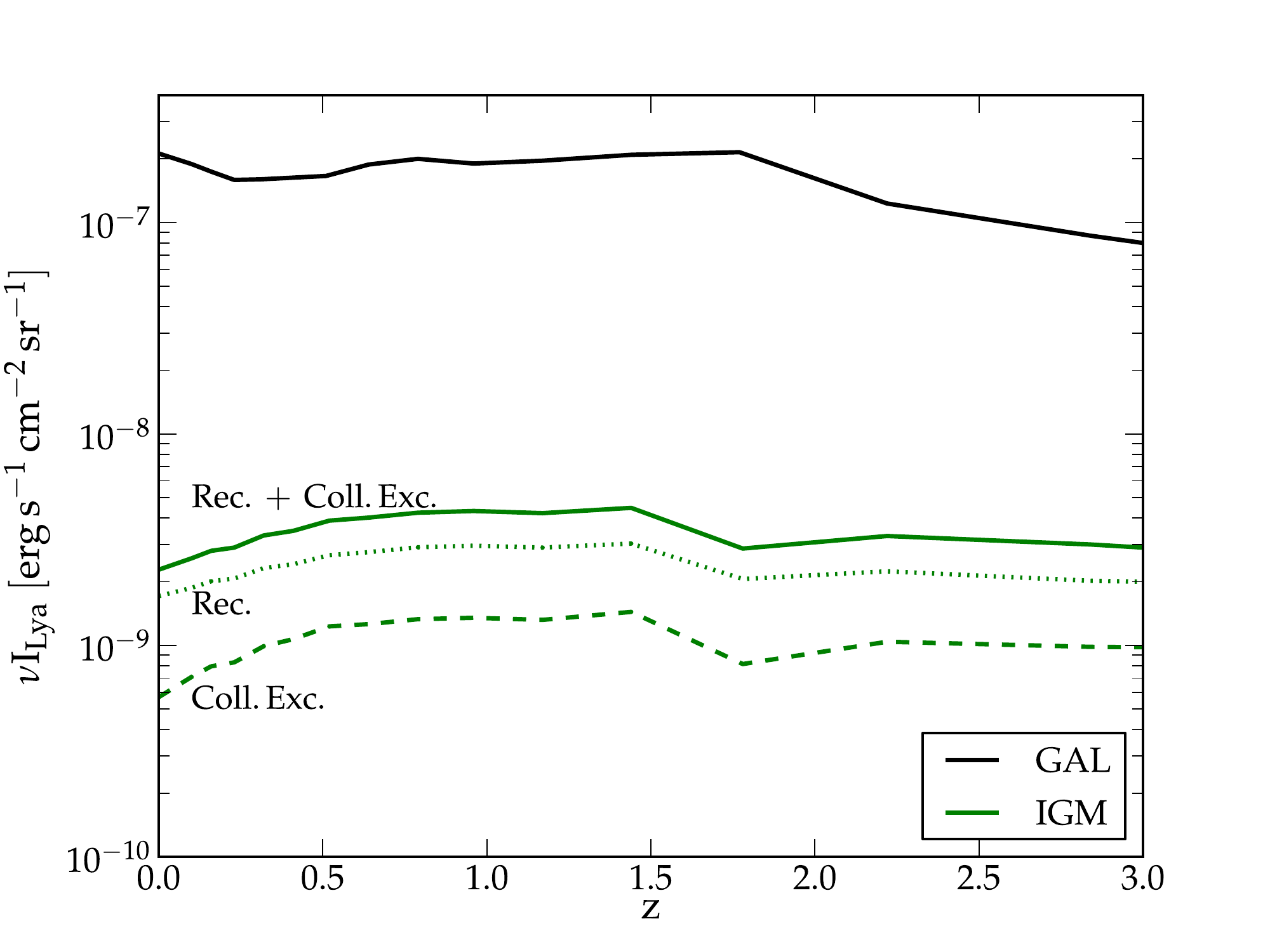}
\caption{Lyman alpha intensity from galaxies and the IGM as a function of redshift obtained from the simulations.}

\label{fig:Inu_z}
\vspace{6pt}
\end{centering}

\end{figure} 
Figure \ref{fig:Teo_lya} shows the theoretical estimate of the \lya intensity from the IGM at $z=0.65$, 
as a function of gas temperature. This figure shows that while in cold IGM filaments \lya emission is mainly 
originated in recombinations in the WHIM collisions will be the main source of \lya photons and the \lya intensity
will be higher by a few orders of magnitude.

Figure \ref{fig:Inu_z} shows the same intensity taken from our simulations but as a function of redshift. 
From the simulations, the \lya emission mainly originates in recombinations from gas with a clumping 
factor of $\sim 5$ and $T_{\rm K}\sim(1-2)\times 10^4\, {\rm K}$ and has an intensity of 
$\nu I_{\rm Ly\alpha}(z=0.65)\sim 2\times 10^{-9}\, {\rm erg\, s^{-1}\, cm^{-2}\, sr^{-1}}$.  
Figure \ref{fig:Inu_z} also shows that galaxies are likely to be the dominant source of \lya emission.
Small variations in our estimates of the thermal and ionization state of the gas might lead to 
an increase (by a factor of a few) of the intensity of \lya emission from collisional excitations, but 
will have little impact on the \lya recombination emission. By using simulations we were able to 
determine the ionization and thermal state of the gas for each cell of the simulation. This is an
improvement over the generalized use of a fixed temperature 
or the use of a temperature parameterized by a power law with a fixed exponent (the adiabatic index), 
which is only appropriate for low density gas ($\delta<10$) cooling adiabatically. Also, the current uncertainty 
in the adiabatic index is relatively high ($1.3>\gamma>1.6$) \cite{2000MNRAS.318..817S,2010ApJ...718..199L,2011MNRAS.410.1096B}, 
which results in a considerable change in the intensity and spatial distribution of the \lya emission.

\subsection{Lyman alpha emission from Galaxies}
\label{sec:model_lya_GAL}

In galaxies the main process leading to \lya emission is recombinations. This \lya emissivity can be roughly estimated as:

\be
\dot{N}^{\rm rec}_{\rm Ly\alpha} = A_{\rm He}\, \dot{N}_{\rm ion}\, (1-f_{\rm esc})\, f_{\rm rec}\, f^{\rm Ly\alpha}_{\rm esc}.
\ee
Here, $\dot{N}_{\rm ion}$ is the rate of emission of ionizing photons by the galaxy, $A_{\rm He}$ is a 
correction factor that accounts for the photons spent in the ionization of helium, $f_{\rm rec}$ accounts 
for the fraction of recombinations that result in the emission of a \lya photon; $f^{\rm Ly\alpha}_{\rm esc}$ 
corresponds to the fraction of \lya photons that escape the galaxy into the IGM and $f_{\rm esc}$ is the 
fraction of ionizing photons that escape the galaxy without being absorbed by dust.

We assume $f_{\rm esc}$ to be around 20\% as predicted by \cite{2014MNRAS.440..776Y} using cosmological 
hydrodynamical simulations and a state of the art radiative transfer code. For low redshift studies the \lya
luminosity is usually estimated assuming $f_{\rm esc}=0$, which is a reasonable approximation in most 
cases, since the fraction of radiation escaping galaxies scales inversely with the galaxy dust content 
and so on average decreases towards low redshifts and high masses, which are the systems that are more 
easily observed. Also, the uncertainty in $f_{\rm esc}$ is very high given the very small number
of direct observational constraints and the large variation of the measured value for different lines of sight.  

The amount of \lya photons emitted due to collisional excitation of neutral hydrogen by free electrons is 
proportional to the leftover energy released into the gas during hydrogen ionizations and thus depends on 
the hardness of the stellar spectrum. 
The number of \lya collisional excitations is  

\be
\dot{N}^{\rm exc} _{\rm Ly\alpha} = A_{\rm He}\, \dot{N}_{\rm ion}\, (1-f_{\rm esc})\, f_{\rm exc}\,f^{\rm Ly\alpha}_{\rm esc}\,  E_{\rm exc}/E_{\rm Ly\alpha}\
\ee
where following the formalism described in \citep{1996ApJ...468..462G} and estimating the average 
energy of an ionizing photon using the spectral energy distribution (SED) of galaxies from the 
\cite{2005MNRAS.362..799M} models, we obtained $E_{\rm exc}\sim 2.14\, {\rm eV}$, for the leftover energy 
per recombination, in a galaxy powered by star formation. 
The emission rate of ionizing photons from galaxies can be estimated from the galaxy SFR as:

\be
\dot{N}_{\rm ion}=Q_{\rm ion} \times {\rm SFR},
\ee
where $Q_{\rm ion}$ is the average number of ionizing photons emitted per solar mass in star formation, which 
at solar 
metallicity and for a Salpeter mass function is given by $2.36\times 10^{60}\, {\rm M}^{-1}_{\odot}$ \citep{2012ApJ...747..100S}.
The resulting intensity of \lya emission from galaxies is

\ba
I^{\rm GAL}_{\rm Ly\alpha}&=&E_{\rm Ly\alpha}\dot{N}_{\rm Ly\alpha}\\ \nonumber
&\approx&  1.55 \times 10^{\rm 42} \left(1-f_{\rm esc}\right)f_{\rm esc}^{\rm Ly\alpha}\frac{\rm SFR}{{\rm M}_\odot\; {\rm yr}^{-1}}{\rm erg\, s^{-1}}.
\ea 

We assume that the fraction of \lya photons which are not absorbed by dust in the ISM is of the order 
of $f^{\rm Ly\alpha}_{\rm esc}=0.3$, which is the average value for $z<3$ also obtained by 
\citep{2014MNRAS.440..776Y}. This is the appropriate 
value for intensity mapping studies, since it accounts for the fraction of \lya photons that escape from 
galaxies and does not account for scattering or absorption of \lya photons in the IGM, since these processes 
conserve the number of photons and therefore the \lya intensity signal as measured by an intensity mapping experiment. 
The much smaller value usually stated in observational studies refers to the effective 
escape fraction of \lya photons, which basically considers that \lya photons are lost 
when scattered out of the line of sight, since these photons will no longer be detected by an observation with a 
small field of view aiming to resolve an individual source \citep{2011ApJ...730....8H,2013MNRAS.435.3333D}. 

We use the halo finder in the Simfast21 code described in \citep{2010MNRAS.406.2421S,2012arXiv1205.1493S}
to extract the DM halos from the N-body simulation. We then estimate their associated star formation 
rate and finally their \lya emission. 
The halos mass is converted into a star formation rate using the relation between these quantities found in the 
\citet{2011MNRAS.413..101G} and \citet{2007MNRAS.375....2D} galaxy catalogs obtained 
by post processing, respectively, the outputs from the 
Millennium II \citep{2009MNRAS.398.1150B} and Millennium I \citep{2005Natur.435..629S} dark matter simulations.
For the mass range available, the parameterizations of the SFR versus halo mass relations were adjusted to fit the 
observationally based constraints from \cite{2015MNRAS.454.2258P} and the simulations 
from \cite{2013ApJ...770...57B}. However, we note that this correction had a small impact in the final results.

\begin{figure*}
\vspace{-10pt}
\centerline{
\resizebox{!}{!}{\includegraphics[scale = 0.50]{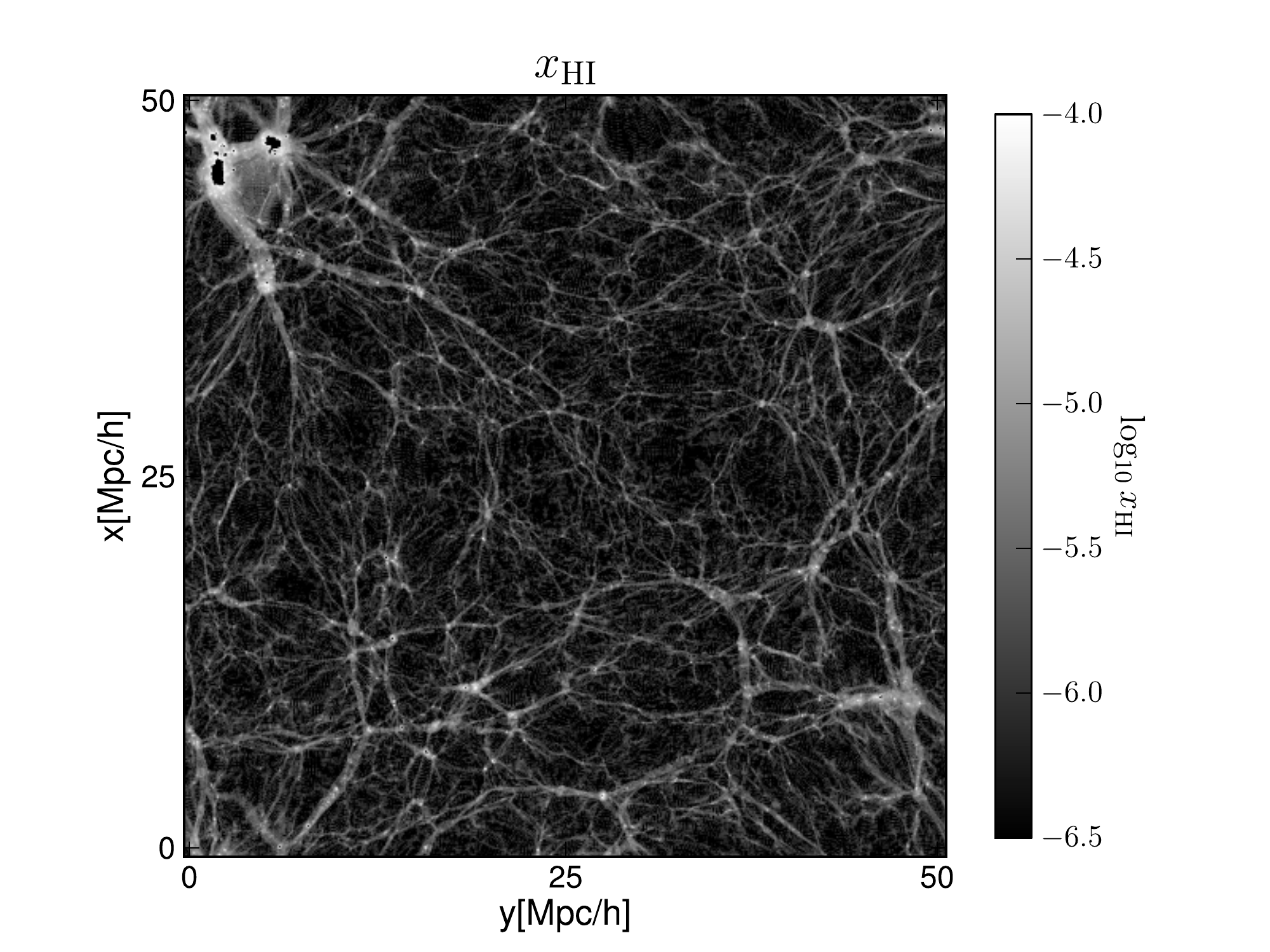}}
\hspace{-28pt}
\resizebox{!}{!}{\includegraphics[scale = 0.50]{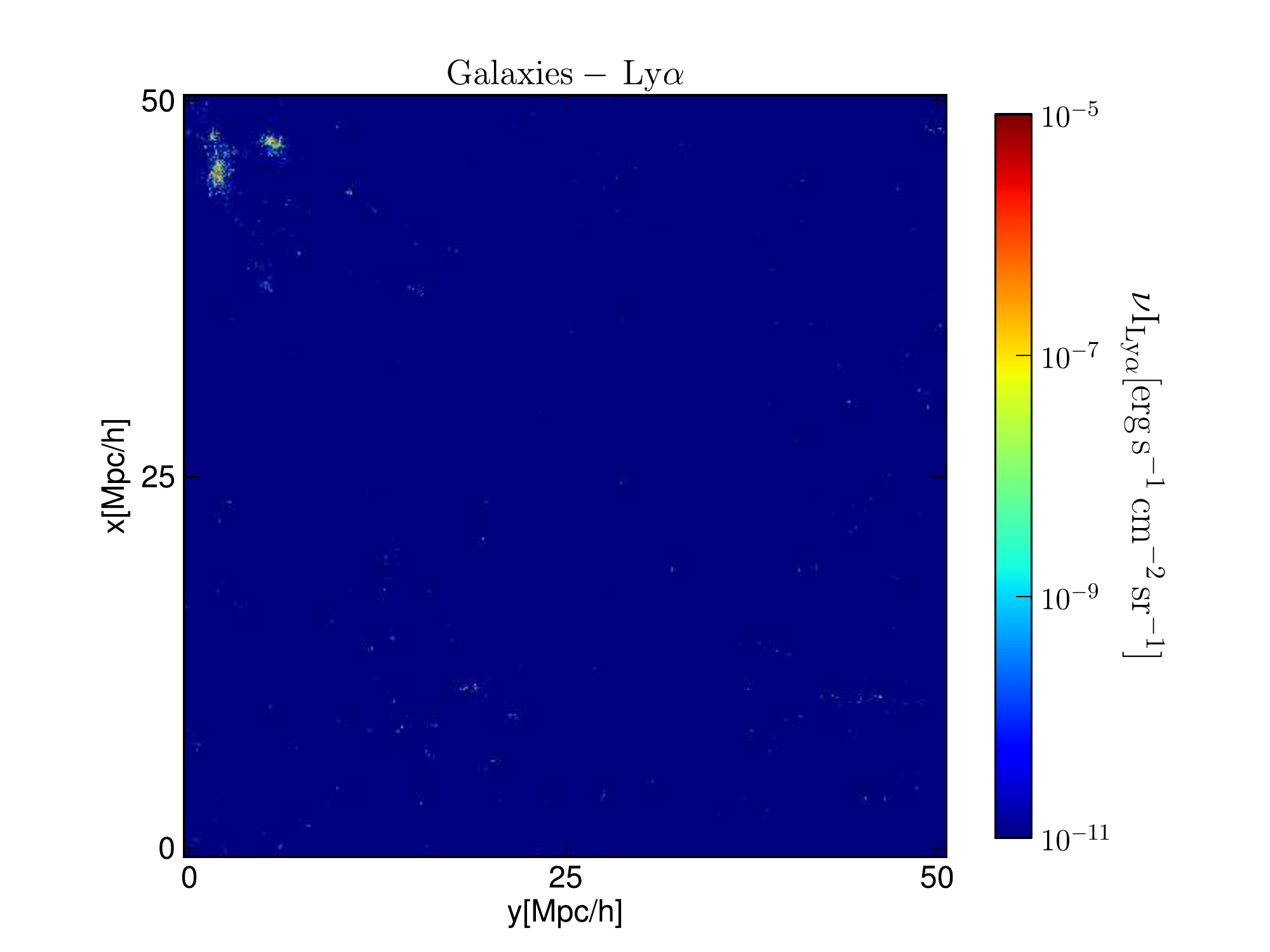}}
}
\vspace{-5pt}
\centerline{
\resizebox{!}{!}{\includegraphics[scale = 0.50]{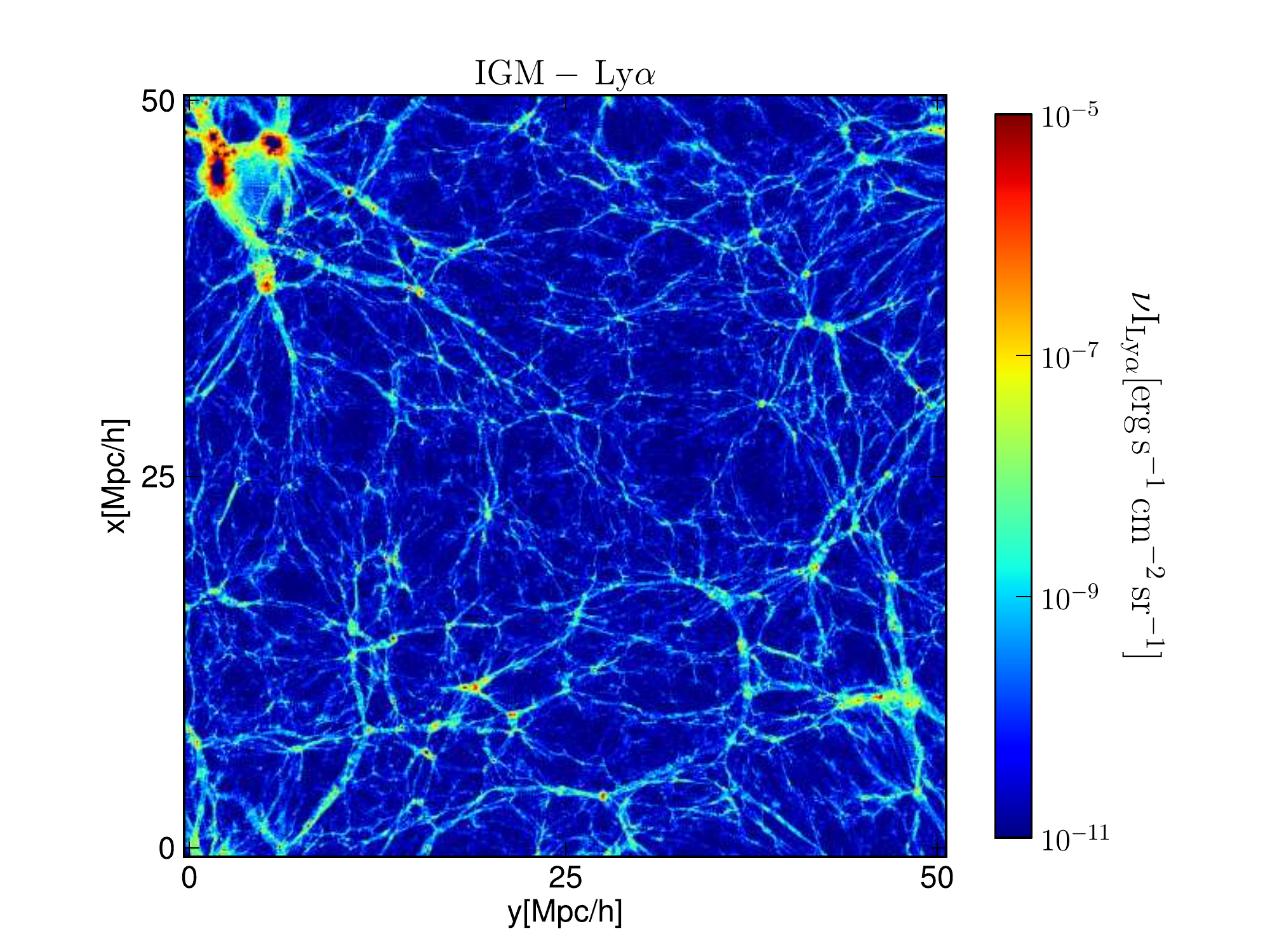}}
\hspace{-28pt}
\includegraphics[scale = 0.50]{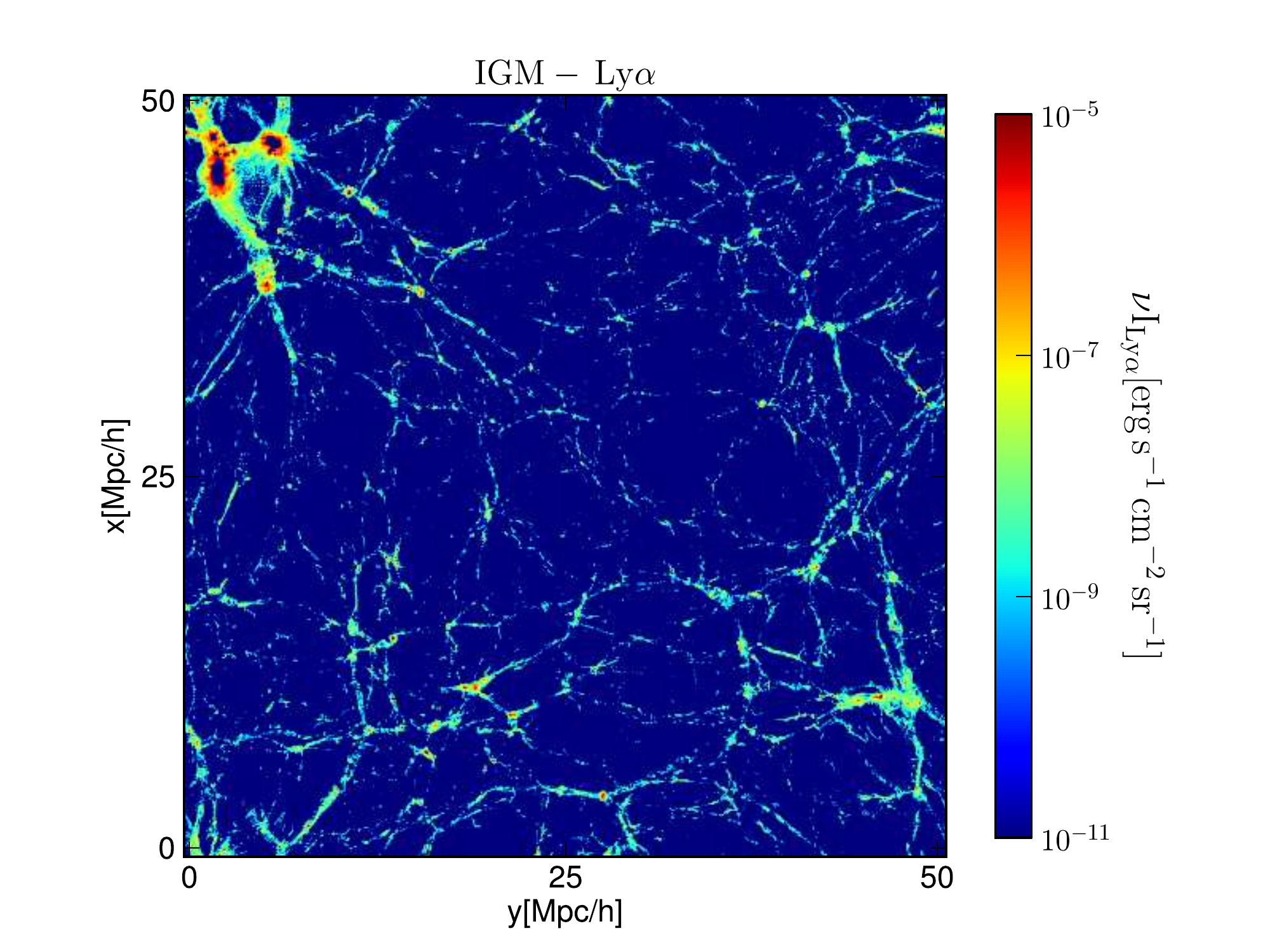} 
}
\vspace{-5pt}
\centerline{
\includegraphics[scale = 0.50]{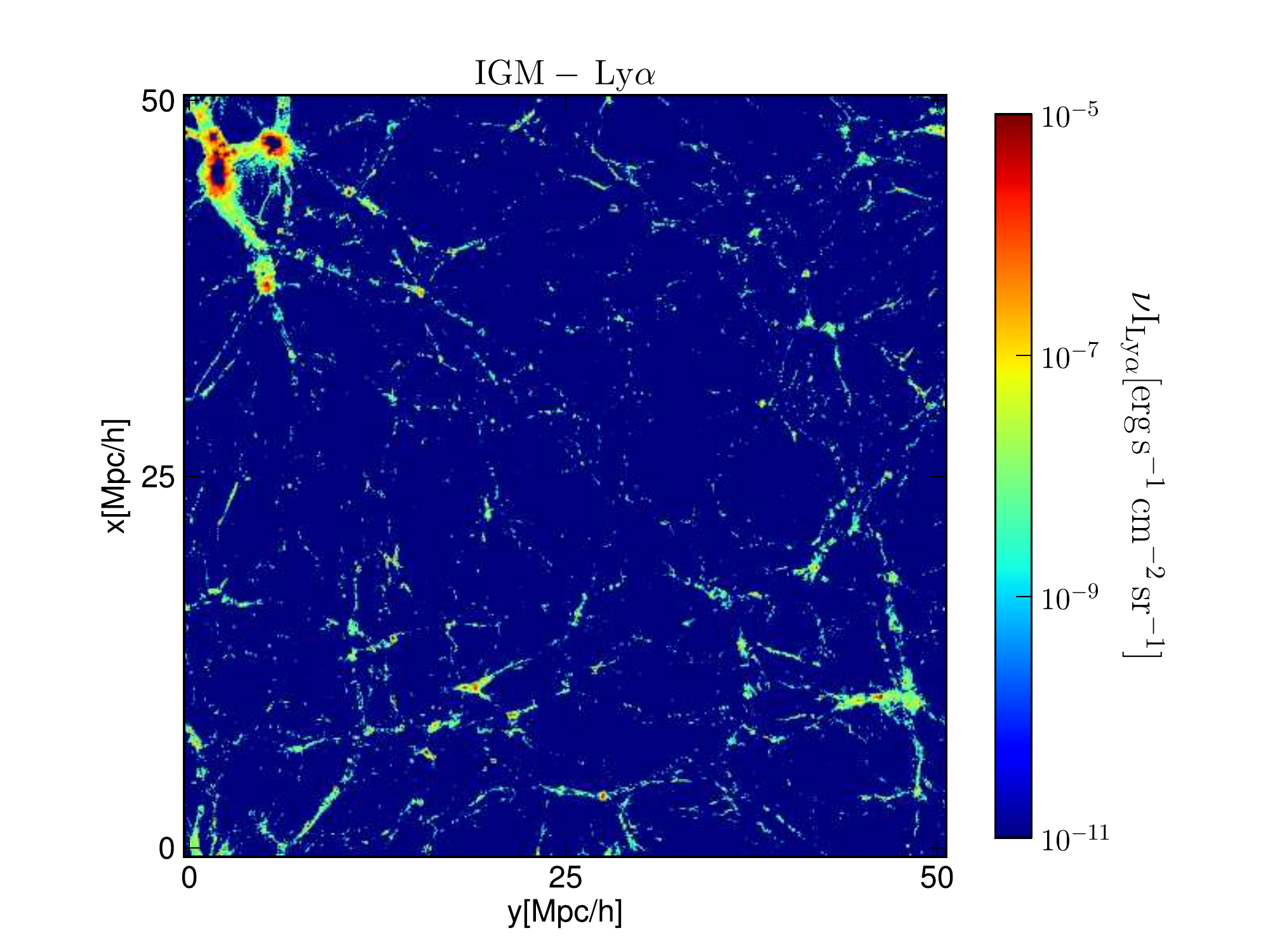} 
\hspace{-28pt}
\includegraphics[scale = 0.50]{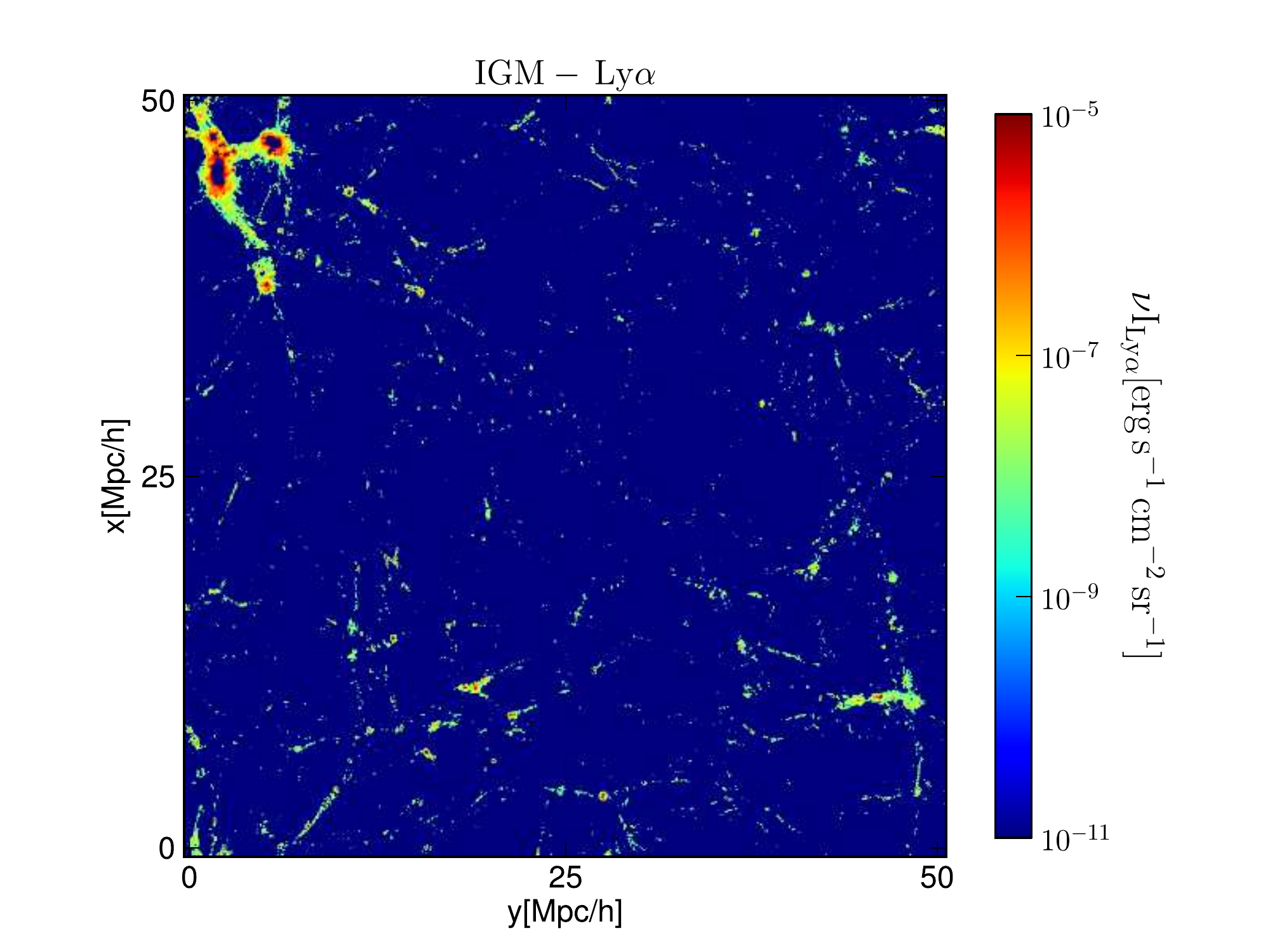} 
}
\caption{\label{fig:4} Simulated maps at ${\rm z=0.65}$. Left upper panel: HI fraction in the IGM. 
Right upper panel: Lyman alpha intensity from galaxies. Except in this panel, galaxies are 
presented as dark regions, since we start by ignoring the properties from these regions.
Middle and lower panels show the intensity of Lyman alpha emission from recombinations and collisional excitations in the IGM. 
These panels from top to down and from left to right assume the full \lya signal, and the \lya signal above an intensity cut of 
${\rm (0.58,1.46,3.66) \times  10^{-9}\, erg\, s^{-1}\, cm^{-2}\, sr^{-1}}$, respectively.
These intensity cuts correspond to flux sensitivities of ${\rm (39, 38 , 37)}$ ${\rm mag/arcs^{2}}$ in the UV band, respectively.}
\label{Fig:maps}
\end{figure*}

\section{Lyman alpha intensity maps and power spectrum}
\label{sec:results}

In the first part of this section the main results from the simulations, the maps of \lya emission, are 
presented assuming experimental setups with different sensitivities. These maps can therefore be used to 
determine the experimental setup required to successfully detect the IGM \lya signal.
We also present a power spectrum analysis of the simulation results. 

\subsection{Lyman alpha intensity results} 
\label{sec:lya_maps}

Figure \ref{Fig:maps} shows the intensity fluctuations in \lya emission from  
galaxies and from the IGM, and how they correlate with the peaks in the HI density.
This figure illustrates how an intensity mapping experiment targeting IGM emission would be able to trace many more 
structures than an instrument targeting only galaxies.
The bottom four panels in Figure \ref{Fig:maps} show the structures observed by 
experiments with decreasing sensitivities. From these panels we can predict that observing the 
main filamentary structures in the IGM through their \lya emission requires
an experiment with a sensitivity of the order of ${\rm \nu I_{\rm Ly\alpha}= 10^{-9}\, erg\, s^{-1}\, cm^{-2}\, sr^{-1}}$.
This intensity corresponds approximately to the intensity of \lya emission from IGM filaments 
assuming the \citep{2012ApJ...746..125H} UV background model. Given the uncertainties in this model, 
we multiplied this background by a factor of two and found that the \lya intensity would only increase by about 20\%,
which would have a minimum impact on our results.

This intensity can be converted to relative magnitudes using the formula from \cite{1983ApJ...266..713O} given by:
\be
 m\, =\, -\, 48.6\, -\, 2.5\, {\rm log}\, f_{\nu}.
\ee
with magnitudes in units of ${\rm mag/arcsec^2}$ and fluxes in units of ${\rm erg\, s^{-1}\, cm^{-2}\, Hz^{-1}\, arcsec^{-2}}$.
Therefore, a surface brightness of ${\rm 37\, mag/arcsec^{2}}$ in the UV band corresponds to a \lya intensity 
of $\nu I_{\rm Ly\alpha}={\rm 3.66\times 10^{-9}\, erg\, s^{-1}\, cm^{-2}\, sr^{-1}}$.
 
\begin{figure}
\begin{centering}  
\includegraphics[angle=0,width=0.51\textwidth]{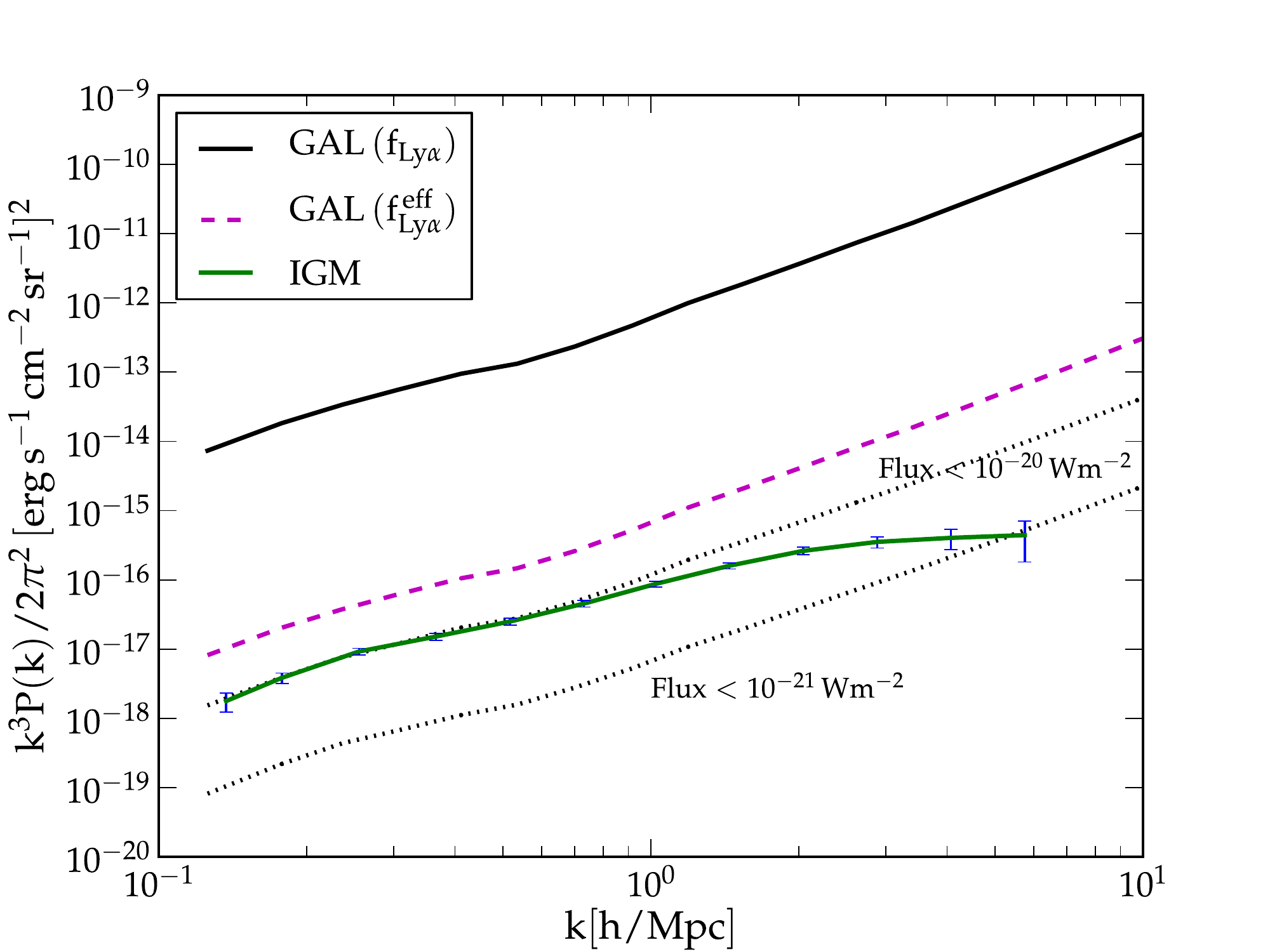}
\caption{Power spectra of \lya emission from galaxies and from the IGM at $z=0.65$. The black 
lines account only for dust absorption of \lya photons in the galaxies ISM while 
the magenta dashed line also accounts for scattering/absorption of \lya photons in the IGM.
The black dotted lines assume only emission from galaxies with \lya fluxes below $10^{-20}\, {\rm W\, m^{-2}}$ 
and $10^{-21}\, {\rm W\, m^{-2}}$. The error bars correspond to the experimental setup of the fiducial instrument described in Section \ref{subsec:IM_lya}.}
\vspace{3 mm}
\label{fig:ps_z065}
\end{centering}
\end{figure} 

\begin{figure}
\begin{centering}  
\includegraphics[angle=0,width=0.51\textwidth]{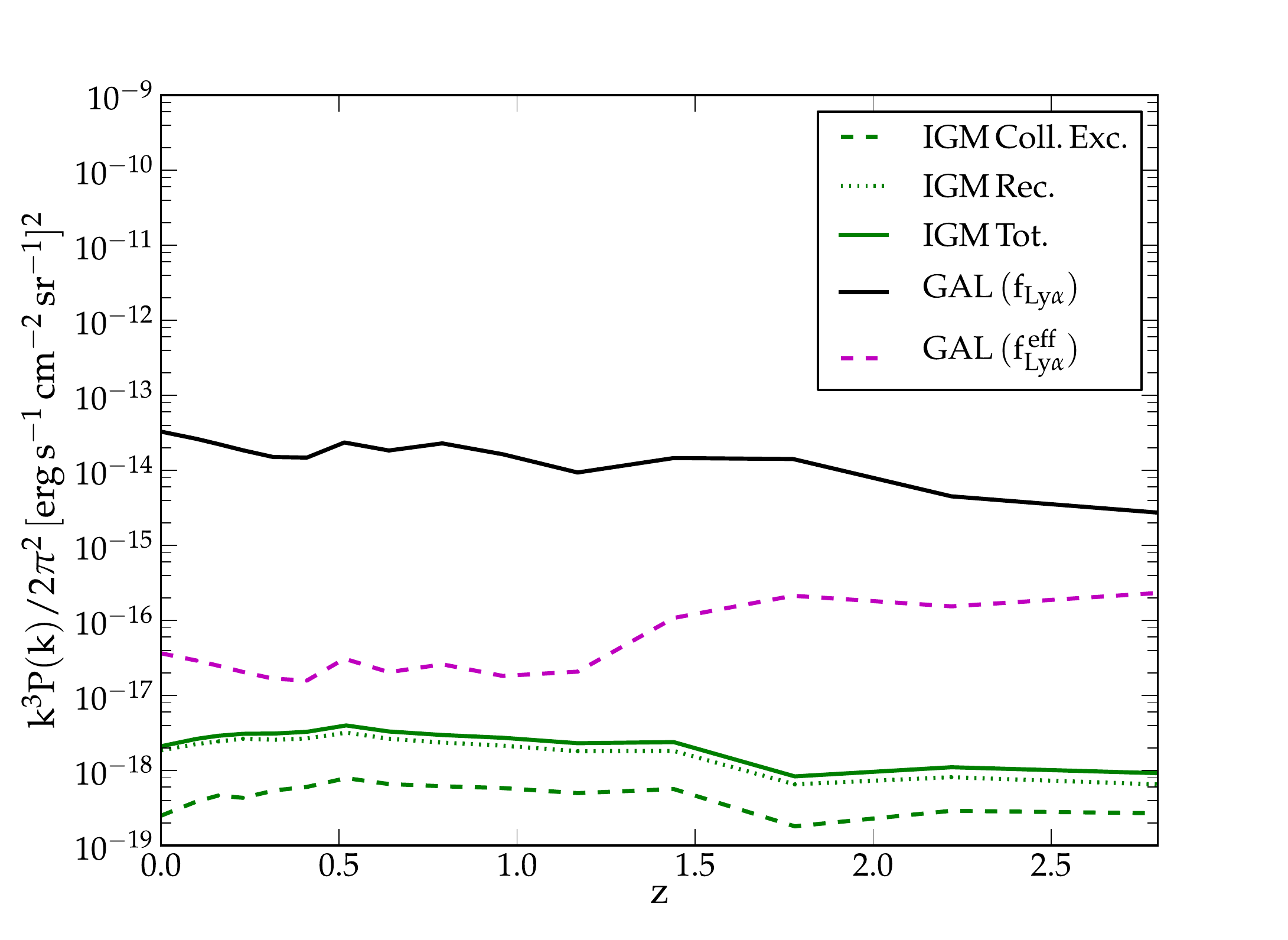}
\caption{Power spectra of \lya emission from galaxies and from the IGM at ${\rm k=0.177\, h/Mpc}$.}
\label{fig:ps_k0.177}
\end{centering}
\end{figure} 

Figure \ref{fig:ps_z065} shows that, at $z=0.65$, the predicted power spectra of \lya emission from galaxies 
is orders 
of magnitude higher than that of emission from the IGM.
Furthermore, Figure \ref{fig:ps_k0.177} shows that this difference is maintained throughout 
the relevant redshift range.
Since the objective of this study is the detection of \lya emission from the IGM we treat \lya emission 
from galaxies as a foreground and determined what fraction of the observational maps need to be masked 
in order to successfully extract the IGM signal.
From our simulations we estimate that, for this to be possible, we need to mask galaxies with \lya fluxes 
above $(10^{-20}-10^{-21}\, {\rm W\, m^{-2}})$, which corresponds to a number density of masked 
galaxies of $\sim  0.11-0.22\, {\rm Mpc^{-3}}$, assuming the experimental setup proposed in Section \ref{subsec:IM_lya}. 
In Figure \ref{fig:ps_z065} we show the contamination power spectra before and after the masking technique is applied. 
As can be observed in this figure, the galaxy power spectrum suffers from a large shot noise contribution at $k\gtrsim 0.06$, 
which makes it easy to distinguish from the power spectrum of IGM fluctuations.
The shape of the observed power spectra can therefore be used to confirm if enough low luminosity galaxies were masked 
and thus if we are indeed detecting the signal from the IGM.

\section{Experiments for detecting UV Lyman Alpha emission}
\label{sec:Experiments}
Currently, there are few experiments probing \lya emission/absorption in the UV regime and these 
missions mainly target strong emission from galaxies or from the WHIM. 
The low surface brightness emission at UV wavelengths is, however, essential to probe galaxies at the 
epoch of peak SFR activity ($z \sim 2$) which is pushing the sensitivity of UV instruments to increasingly 
lower fluxes and will bring them to the level necessary to also probe \lya emission from IGM filaments.
Probing the low surface brightness universe is limited by systematics, sky variability, straylight, 
flat field accuracy and extended PSF wings and only now, with the advent of technological advances, is this 
goal becoming possible.

Examples of current or proposed experiments that propose to probe IGM emission include the following 
missions. The Faint Intergalactic Redshifted Emission Ballon (FIREBall2) is able to measure 
\lya, OVI and CIV emission from the WHIM at 205 ${\rm nm}$ \citep{2010AAS...21531506T}, and its sucessor, 
the proposed Imaging Space Telescope for Origins Survey (ISTOS) plans on targeting \lya emission 
from ${\rm z\sim 0.5-1.2}$. There are also several orbital sounding rockets carrying integral field 
spectrometers able to detect \lya emission from galaxies and from the CGM in the redshift range 
${\rm (z\sim 0 - 1.5)}$ \citep{2015arXiv151200881F}. The Cosmic Origins Spectrograph 
(COS) on board of the Hubble Space Telescope also targets \lya emission from galaxies and from 
the CGM in the redshift range ${\rm (z\sim 0 - 1.5)}$ \citep{1998SPIE.3356..361M,2003SPIE.4854...72G}.

These experiments will, however, not be able to detect the faint population of \lya emitters or the cold IGM.
On the other hand, the recently proposed Messier satellite plans to achieve unprecedented surface brightness 
levels of ${\rm 37\, mag/arcsec^2}$ in the UV, which is enough to probe emission from faint galaxies, from the 
WHIM and even from the cold IGM. This satellite will carry a photometer with a very high spatial resolution 
but with mostly broadband frequency filters. The main sensitivity of the satellite will be at ${\rm z\sim 0.65}$, since 
it will have a narrow band filter  centered at ${\rm 200\, nm}$.
The proposed frequency band of ${\rm \Delta \lambda \sim 50\, nm}$, however, is too large to identify IGM filaments, 
although it will be sensitive to the overall emission from several filaments. We note that the width of IGM 
filaments should be of the order of ${\rm 0.5\, Mpc/h}$ \citep{2014MNRAS.438.3465T}. 

\subsection{Experimental setup for Lyman Alpha emission Intensity Mapping}
\label{subsec:IM_lya}

In this section we describe the experimental setup of a UV satellite capable 
of detecting and mapping \lya emission from IGM filaments and thus also able to probe the low surface brightness universe. 
The setup of the described detector is appropriate, not only to detect this signal, 
but also to study the statistical properties and the spatial distribution of IGM filaments.
If we just aim at detecting the filaments, an experiment with a smaller resolution would suffice.

We take a UV sensitivity of ${\rm 37\, mag/arcsec^2}$ as the minimum surface brightness level 
required to detect most ($>90 \%$) of the emission from faint filaments in the IGM at $z\, =\, 0.65$, 
as estimated in Section \ref{sec:lya_maps}. This UV magnitude corresponds to the detection of \lya emission
above ${\rm \nu I_{Ly\alpha}\sim 3.7 \times 10^{-9}\, erg\, s^{-1}\, cm^{-2}\, sr^{-1}}$.

A lower 
sensitivity would still allow the detection of many IGM filaments. However, since we propose to analyze 
these filaments at a statistical level, using their intensity and power spectra, the more filaments we can 
detect the more we can say about their properties.
  
Given the length of up to ${\rm 60\, Mpc/h}$ of the filaments recently identified by \cite{2014MNRAS.438.3465T} 
in the SDSS public galaxies catalog, we assume the minimum area to be observed by the proposed experiment to 
be of ${\rm 60\, Mpc/h}$, which at $z \sim 0.65$ requires at least a $2 \degree \times 2 \degree$ observational field of view.
The ideal spatial resolution necessary to identify the filaments is of the order of ${\rm 1\, Mpc}$. Lower resolutions 
would however still allow for the statistical detection of this 
signal, but would make the masking of foreground emission very difficult. This spatial resolution 
corresponds to a frequency resolution of R=1000 and an angular resolution of 85\arcsec. A telescope with an 80 cm 
aperture and 200$\times$ 200 elements could detect this signal, with high precision, in 3000 hours, as shown in 
Figure \ref{fig:ps_z065}.

Since we only want to detect emission from the IGM, we will need to mask pixels that contain emission from bright 
galaxies, which for the previously referred resolution would result in ${\rm 10\, -\, 20 \%}$ of the pixels being 
masked. This would not erase much of the information about the filamentary structures (see Sections 
\ref{sec:lya_maps} and \ref{sec:Line_contamination} for 
calculations of the number density of pixels that need to be masked).
A higher spatial resolution would, however, easily reduce this masking percentage even more.

Due to the nature of the \lya line, its FWHM (full-width at half-maximum) 
will limit the maximum possible resolution of the experiment, or require that galaxy 
emission is masked from several pixels.

We note that a masking of 20\% of the pixels would still allow the
recovering of the intensity and power spectra of Lyman alpha emission from the cold IGM, since most 
of the masked intergalactic gas would correspond to warm/hot gas in the circumgalactic medium.

For the masking of bright galaxies, a complementary experiment would be required. Ideally this can be 
done with the same spectrometer required to detect the map of the emission from IGM filaments. This is a reasonable
assumption, since the only difference necessary for the same experimental setup to be able to detect these galaxies
would be a higher spatial resolution of the order of 1\arcsec per pixel.  This could be achieved with the planned Messier 
experiment if it had a larger spectral resolution, or with the Multi Unit Spectroscopic Explorer (MUSE) at the VLT if it 
had a larger field of view and could be extended 
to lower frequencies \citep{2014Msngr.157...13B}. Another promising experiment is the World Space Observatory - Ultraviolet (WSO-UV) space telescope,
which should be launched in 2021 and aims to detect UV emission in the frequency range corresponding to \lya emission from $z\sim 0 - 1.5$ 
\citep{Boyarchuk2016}. The proposed (WSO-UV) telescope will has the required 
resolution and sensitivity for detecting both \lya emission from IGM filaments and from galaxies up to the required flux cut. However, the
field of view of this telescope is only $30\arcmin \times 30\arcmin$ and so not big enough to study IGM filaments at a statistical level.
  
The technology to detect \lya emission from the contaminant galaxies thus already exists, but currently none of the proposed experiments 
meets all the requirements. Alternative, these galaxies 
can be detected through their H$\alpha$ emission, since this is a good probe of the galaxies 
intrinsic \lya emission. For the relevant frequency range, there are photometric and spectroscopic instruments 
that can detect galaxies through their H$\alpha$ down to one order of magnitude above the necessary flux limit, 
or down to $10^{-19}\, {\rm W\, m^{-2}}$ in \lya emission \citep{2013MNRAS.428.1128S}.

\section{Foregrounds in Lyman alpha intensity maps}
\label{sec:Foregrounds}

Lyman alpha intensity maps will be contaminated by both line and continuum foregrounds. 
The contamination by line foregrounds can be considerably reduced by masking the pixels contaminated 
by the most luminous foreground sources, similarly to what was done in \cite{2014ApJ...785...72G} for 
high redshift \lya emission. We discuss the line contamination in \lya intensity maps at $z<2$ in 
Section \ref{sec:Line_contamination}.

Continuum foregrounds include free-free, free-bound, two photon and UV emission from stars and 
active galactic nuclei.
Most continuum emission can be fitted out of observational maps, since this continuum evolves with frequency 
more smoothly than the target line emission, which varies according to the emitting structures along the line of sight.

In intensity mapping studies, another important source of radiation is the UV background radiation, which
redshifts into the \lya frequency at the target redshift \citep{2013ApJ...763..132S,2014ApJ...786..111P}.
Although the photons in the UV background are also continuum emission, if they redshift into the \lya frequency in 
a region with a high density of neutral hydrogen, they will scatter in the gas and 
get re-emitted in a random direction. Therefore, this signal will correlate with the structures in the IGM and will no 
longer evolve smoothly with frequency. At the low redshifts we are interested in, the probability 
of a UV photon encountering a region with neutral hydrogen as it redshifts into one of the Lyman-n 
lines is considerably small, taking into account the highly ionized state of the gas and the small 
natural width of the \lya line. We estimate the intensity of this emission as

\begin{eqnarray}\label{E:Icont}
I_{\rm diff\,cont}^{\rm Ly\alpha}(z) &=& \frac{h\, \nu_{\rm Ly\alpha}}{4\pi}\sum_{n=2}^{\infty}f_{\rm rec}(n)\int_z^\infty dz'\,\frac{c}{(1+z')\, H(z')}\nonumber\\
&&\times\, {\rm SFRD}(z')\, \epsilon(\nu_n')\, P_{\rm abs}(n,z)\nonumber\\
&&\times\prod_{n'=n+1}^{n_{\rm max}}\{1-P_{\rm abs}[n',z_{n'}(z,n)]\}\, .
\end{eqnarray}

is the taken from \cite{2005ApJ...620..553L} and modified by \cite{2014ApJ...786..111P} to account for the case where 
a fraction of the IGM is ionized. 
Here $\epsilon(\nu_n)$ is the emissivity rate of a Lyn photon and $f_{\rm rec(n)}$ accounts for the probability 
of a Lyn photon to cascade down to a \lya photon.
By using appropriate parameters for the absorption probability ($P_{\rm abs}$) in 
the low redshift, highly ionized gas, the resulting scattered emission is extremely low. Our
result differs from the high scattering probability found by \cite{2014ApJ...786..111P},
since we determined the optical depth for \lya absorption with a high resolution simulation instead of 
using a simplified formula to estimate the probability of neutral hydrogen clumps in the IGM which is 
not appropriate for the very ionized gas at $z\, <\, 2$ where the IGM gas is highly ionized by the strong UV 
background. We found that, even at $z\, =\, 0$, where this scattered \lya emission
should be the highest, our most optimistic values for this emission 
are of the order of ${\rm  I_{Ly\alpha}}(z\sim0) = {\rm 10^{-29}\, erg\, s^{-1}\, cm^{-2}\, sr^{-1}\, Hz^{-1}}$. On 
the other hand, the continuum background at the \lya frequency will be several orders of magnitude above 
the proposed signal, see Section \ref{sec:Cont_contamination}.
Therefore, we can consider the UV background as a continuum foreground 
that evolves smoothly with frequency. 

The existence of a strong UV photon source, such as an active galactic nucleus, in the background of a dense 
gas filament will also produce extended \lya emission which, depending on its strength, might leave 
a signature in \lya intensity maps. The number density of active galactic nuclei is, however, relatively 
small. Therefore, considering the size of the observational pixel proposed, they can be easily masked without 
erasing a meaningful fraction of the \lya signal from the IGM.

An additional complication to the detection of IGM \lya emission arises from \lya extended emission around 
galaxies, typically known as \lya blobs. The source of power for these large nebulae can vary between energy released 
during gravitational collapse, star burst activity or active galactic nuclei emission. 
The failure of GALEX to observe \lya blobs at $z=0.8$ indicates that these structures are restricted to high redshifts 
and so they are not relevant foregrounds for intensity mapping studies at low redshift. 

Recent observations of \lya blobs indicate that these phenomena occur in high density 
regions and that they are mainly powered by star formation \citep{2015A&A...581A.132A,2016arXiv160100682A}. 
Given their proximity to star forming regions and AGN, most \lya blobs will be removed from the observational 
maps during the galaxy masking. We note that, although the spatial extension of these sources can reach up to 100 
kpc, that size is well below the proposed pixel resolution \citep{2013ApJ...773..151Y}. 

The \lya blobs remaining in the 
observational maps will therefore be subdominant compared to emission from the IGM.

\subsection{Line contamination}
\label{sec:Line_contamination}

For the redshift range $z\sim 0 - 3$, the major line contaminant in \lya intensity maps is 
the ${\rm [OII]\, 372.2\, nm}$ line emission
from redshifts $z\sim0 - 0.3$, which is observed at the same frequency as \lya from redshifts $z\sim 2.07 - 3$.

The OII line luminosity can be estimated with the relation 
\be
L_{\rm [OII]}\, ({\rm erg\, s^{-1}})=7.1 \times 10^{40}\, {\rm SFR\, (M_{\odot}\, yr^{-1})}
\ee
from \cite{1998ApJ...498..541K} and with the SFR parameterization described in Section \ref{sec:Lya_emission}.
The resulting OII intensity for $z\sim0.046$ that will contaminate \lya intensity maps at $z\sim2.2$ is
$\sim1.58\times10^{-21}\, {\rm erg\, s^{-1}\, cm^{-2}\, sr^{-1}}$.
We estimated the contamination power spectrum from OII emission scaled to the redshift at which \lya 
is emitted to be approximately two orders of magnitude above 
the \lya signal power spectrum at large scales. The required flux cut in the OII emitting galaxies 
needed to decrease the contamination power spectrum 1 order of magnitude below signal power spectrum is 
$L_{\rm [OII]}\sim1.7\, {\rm erg\, s^{-1}}$. This flux cut corresponds to masking only around 
$0.09\, {\rm Mpc^{-3}}$ galaxies.
Note that, in the redshift range $2.2\lesssim z \lesssim2.06$, where the projection of the 
contamination power spectra to the redshift of \lya emission will be particularly high, and so 
it will be difficult to mask maps in this redshift range. According to this study, the masking percentages 
required for maps at $z>2.2$ are smaller than the ones presented at $z=2.2$.

When it is not possible to spectroscopically observe the contaminant OII galaxies, 
they can be photometrically observed and distinguished from \lya emitting galaxies due to the 
much smaller EW of the OII line \citep{2014IAUS..306..365A}. 

\subsection{Continuum contamination} 
\label{sec:Cont_contamination}

Continuum stellar photons emitted with frequencies within the \lya equivalent width 
($\sim 1 {\rm \mathring{A}}$) should scatter within the ISM and eventually get 
re-emitted out of the galaxy as \lya photons \citep{2013MNRAS.428.1366J}. This emission 
is very weak compared to the nebular \lya emission and can therefore be safely neglected. 

However, continuum stellar emission below and above the \lya frequency will, contaminate \lya intensity 
maps.
We estimate this contamination by assuming that the number of photons emitted 
by a galaxy within the relevant frequency range follows the spectrum of a 
black body with an average temperature of $\sim2.2\times{\rm 10^4\, K}$. We estimate, the black body 
temperature at each emitted frequency by averaging over a Salpeter stellar mass function, the temperature of the stars weighted by the 
lifetime of their nuclear phase and by their relative emissivity at the relevant frequency range.

The galaxy spectrum, amplitude was set, assuming that the average relation between the galaxy 
SFR and its luminosity in the range $1500\, {\rm \mathring{A}}<\lambda<2800\, {\rm \mathring{A}}$
follows the relation
\be
\frac{{\rm SFRD}(z)}{[{\rm M_{\odot}\, yr^{-1}\, Mpc^{-3}}]}=1.05\times 10^{-28} \frac{\rho_{\rm UV}}{[{\rm erg\, s^{-1}\, Mpc^{-3}\, Hz^{-1}}]},
\label{eq:SFRD_LUV}
\ee
where $\rho_{\rm UV}$ is the galaxies luminosity density \citep{2012ApJ...746..125H}. The continuum stellar 
intensity at the observed frequency is  
\ba
I^{\rm stellar}_{\nu_o}(z_{\rm Ly\alpha})&=& A \int_{\rm \nu_{\rm min}}^{\rm \nu_{\rm max}} d\nu \frac{h \nu^3}{e^{\rm h\nu/k_B T_K}-1}\nonumber \\  
&\times& \frac{{\rm SFRD}(z)y(z)}{4\pi(1+z)^2}f^{\rm UV}_{\rm esc},
\label{eq:Qstellar_cont_lya}
\ea
where $A\, =\, 4.19\times10^{9}$ sets the number of UV photons estimated with the black body approximation to 
follow the relation presented in equation \ref{eq:SFRD_LUV}. Here, $\nu_o$ is the frequency at which the 
lyman alpha line is observed when emitted from the redshift $z_{\rm Ly\alpha}$, $\nu=\nu_o(1+z)$ is the 
emitted frequency and $f^{\rm UV}_{\rm esc}$ is the fraction of the emitted UV photons that escape the 
galaxy without being absorbed by dust. We integrate the continuum emission from 
$\nu_{\rm min}=\nu_{\rm Ly\alpha}/(1+z_{\rm Ly\alpha})$ to
$\nu_{\rm max}=7.4\times10^{15}\, {\rm Hz}$, which corresponds to photons with energies $30.6\,{\rm eV}$.
For most stars the radiation released above this energy is very small, so our results are not very sensitive to the choice of $\nu_{\rm max}$. 

We parameterize the escape fraction of UV photons below the 
Lyman alpha frequency as $f^{\rm UV}_{\rm esc}=10^{-0.4A(\nu,z)k(\nu)}$ and, following \cite{2012ApJ...746..125H}, 
we take the magnitude of the attenuation to be:

\be 
A(\nu,z)=A_{\rm FUV}\frac{k(\nu)}{k(1500 {\rm \mathring{A}})},
\ee
where  

\begin{equation}
A_{\rm FUV}(z)=\begin{cases} 1 & (0\le z\le 2);\\
2.5 \log [(1+1.5/(z-1)] & (z>2).
\end{cases}
\label{eq:AFUV}
\end{equation}

To scale the attenuation with frequency we use the starburst reddening curve, $k(\nu)$, from \cite{2000ApJ...533..682C}.

We take $f^{\rm UV}_{\rm esc}=0.3$ for emission between the 
\lya and the Lyman limit frequencies and $f^{\rm UV}_{\rm esc}=0.2$ for ionizing UV emission  
(following the study by \cite{2014MNRAS.440..776Y}). Given the uncertainty in the escape fraction of photons 
above the \lya frequency, a lower limit for this intensity can be obtained by setting 
${ z_{\rm max}=z_{\rm Ly\alpha}}$.

Additional free-free, free-bound and two-photon radiation emitted from redshifts below 
the \lya target redshift will also contribute to the ultraviolet background radiation.
Free-free and free-bound emission is, respectively, originated when free electrons 
scatter off 
ions with or without being captured. The intensity of free-free and free-bound radiation can be modeled as:
\be
I^{\rm free}_{\nu_o}(z_{\rm Ly\alpha})= \int_{\rm \nu_{\rm min}}^{\rm \nu_{\rm max}} d\nu\, f_{\rm volume}(z)\frac{\varepsilon_{\nu}}{4\pi (1+z)^2}y(z),
\ee
where $\varepsilon_{\nu}$ is the total volume emissivity (which we estimated as in \citep{2003adu..book.....D}) and $f_{\rm volume}$ 
is the volume fraction of the Universe which contains free electrons and ions.

At the relevant redshift range most of the gas in the IGM is ionized. However, for estimating $f_{\rm volume}$ it is useful
to divide the IGM in three regions which have very different astrophysical properties. The CGM, which is very dense 
($\delta>100$), has a temperature of $T_{\rm K}\sim 10^5\, {\rm K}$ and contains around 5\% of the baryons.  
The IGM gas filaments, with overdensities in the range $1>\delta>180$ has a temperature of $T_{\rm K}\sim 10^4\, {\rm K}$ and 
contains more than 40\% of the baryons. 
Finally, we consider the voids. These are characterized by overdensities of $\delta<1$ and only 
contain a small fraction of the baryons in the Universe. Given the low luminosities involved, it is safe to neglect 
emission from voids.
Therefore, we estimate the volume fraction of the IGM emitting continuum radiation as the volume 
occupied by filaments, which is around 5\% of the total volume in our simulations, in agreement with 
the results from N-body simulations and from the SDSS detected filaments analyzed by \cite{2014MNRAS.438.3465T}. 

To estimate free-free and free-bound emission 
from filaments, it is a good approximation to assume that all of the gas is ionized and that it has a clumping factor 
of $C\sim 5$ (value estimated from our simulations). 
The volume emissivity estimated by \cite{2003adu..book.....D} is given by:
\be
\varepsilon_{\nu}=10^{8.1} C n_{\rm e} n_{\rm p} \gamma_c \frac{e^{-h\nu/k_{\rm B} T_{\rm K}}}{T_{\rm K}^{1/2}}\, {\rm erg\, cm^{-3}\, s^{-1}\, Hz^{-1}}, 
\ee
where $\gamma_c$ is the continuum emission coefficient, including free-free and free-bound emission given, 
in SI units by:
\be
\gamma_c=5.44\times 10^{-46}\left[\bar{g}_{ff}+\Sigma_{n=n^\prime}^{\infty}\frac{x_n e^{x_n}}{n}g_{fb}(n)\right].
\label{eq:gamma}
\ee
Here, $x_{\rm n}=Ry/(k_{\rm B} T_{\rm K} n^2)$, where $k_{\rm B}$ is the Boltzmann constant, n is the level to which the electron 
recombines, $Ry=13.6\, {\rm eV}$ and $\bar{g}_{ff}\approx1.1-1.2$ and $g_{fb}(n)\approx1.05-1.09$ are the 
thermally average Gaunt factors for free-free and free-bound emission, respectively, \citep{1961ApJS....6..167K}. 
The initial level $n^\prime$ is set by the 
condition $cR_{\infty}/n^{\prime 2} <\nu < c R_{\infty}/(n^\prime-1)^2$, where $R_{\infty}=1.1\times10^7\, {\rm m^{-1}}$
is the Rydberg constant. 
In an ionized gas, the number density of protons is $n_p=n_{\rm H}+2n_{\rm He}$ and the number density 
of electrons is $n_e=n_{\rm H}+2n_{\rm He}$. 

The average number densities in the filaments from the simulation are therefore $n_{\rm H}=\frac{0.4}{0.05}(1-Y_p)\bar{n}_b\sim1.49\times 10^{-6}\, {\rm cm^{-3}}$
and $n_{\rm He}=\frac{0.4}{0.05}Y_p\bar{n}_b/4\sim1.23\times 10^{-7}\, {\rm cm^{-3}}$, where $\bar{n}_b$ is 
the baryon number density.
Similarly, during a recombination, there is a probability of two photons being emitted. 

The observed intensity due to 2-photon emission is
\ba
I^{\rm 2-photon}_{\nu_o}(z_{\rm Ly\alpha})&=& C n_{\rm e} n_{\rm p} \alpha_A (1-f_{\rm Ly\alpha})\nonumber \\
&\times& \int_{\rm \nu_{\rm min}}^{\rm \nu_{\rm Ly\alpha}} d\nu \frac{2h\nu}{\nu_{Ly\alpha}}\frac{P\left(\nu/\nu_{Ly\alpha}\right)}{4\pi}\nonumber \\
&\times& (1+z)y(z),
\ea
where P(y)dy is the normalized probability that, in a two photon decay, one of them is in the range $dy=d\nu/\nu_{ly\alpha}$. 
The factor $1-f_{ly\alpha} \approx 1/3$ accounts for the probability of 2-photon emission during a hydrogen recombination. 
This probability was fitted by \cite{2006ApJ...646..703F}, using Table 4 of Brown $\&$ Mathews (1970), as:
\ba
P(y)&=&1.307-2.627(y-0.5)^2+2.563(y-0.5)^4\\ \nonumber
    &-&51.69(y-0.5)^6.
\ea

The last missing contribution to the UV continuum background is emission from active galactic nuclei which we 
modeled using, as a lower limit, the quasar comoving emissivity at 1 ryd from \cite{2007ApJ...654..731H}, fitted 
by \cite{2012ApJ...746..125H} as:
\begin{figure}
\begin{center}  
\hspace{-10pt}
\includegraphics[angle=0,width=0.51\textwidth]{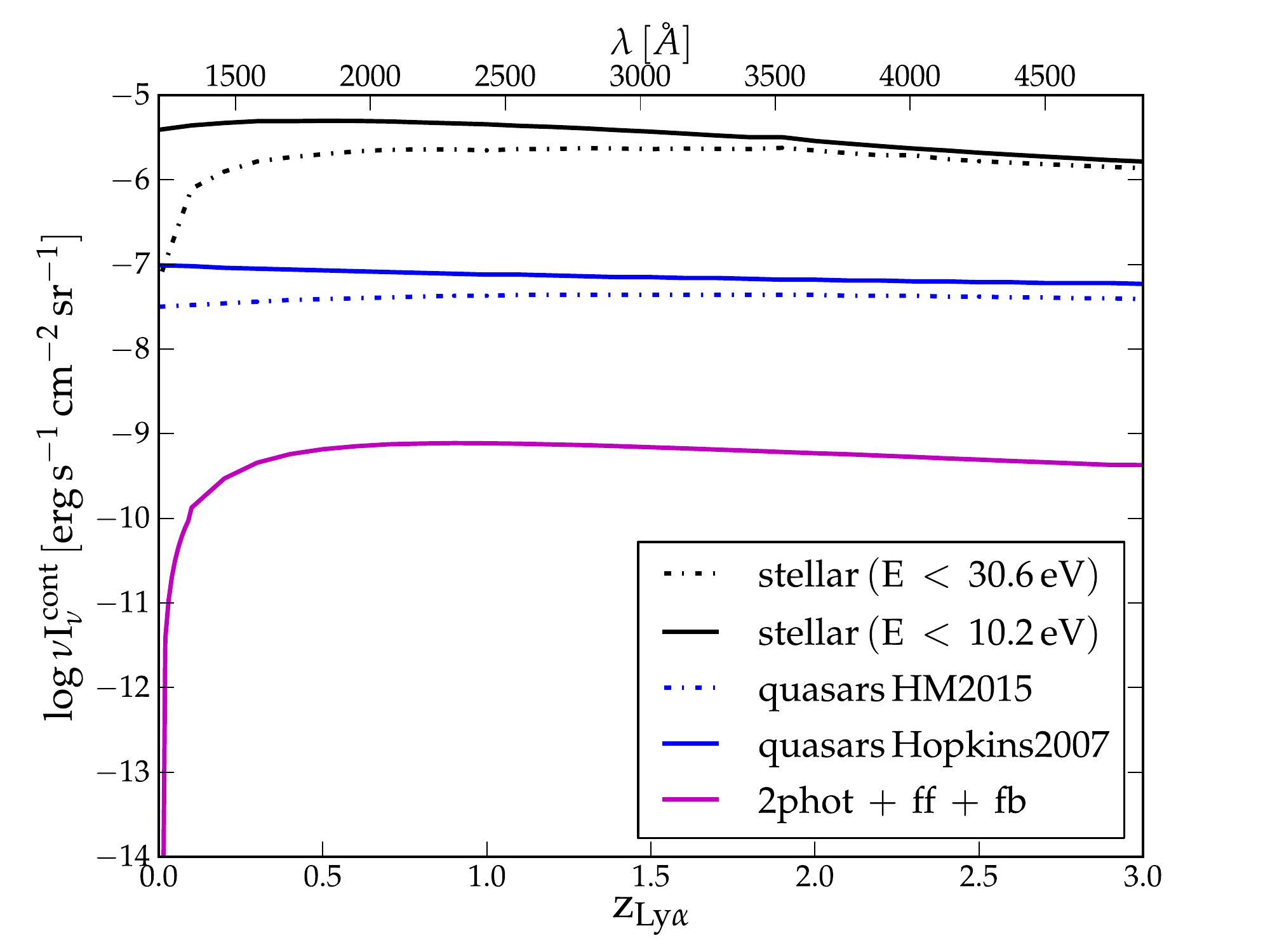}
\caption{Intensity of continuum background contamination in Lyman alpha intensity maps as a function of the 
Lyman alpha emission redshift. The black lines denote the predicted intensity from stellar UV 
emission. The upper black solid line includes UV emission below $30.6\, {\rm eV}$, while the lower black 
solid line only includes UV emission below $10.2\, {\rm eV}$. The blue lines account for quasar emission 
according to the \citet{2007ApJ...654..731H} and the \citet{2015ApJ...813L...8M} models. 
The magenta line includes continuum contamination from 2-photon, free-free and free-bound emission.
}
\label{fig:cont_z}
\end{center}
\end{figure} 
\ba
\frac{\epsilon_{912}(z)}{(1+z)^3}&=&\left(10^{24.6}\, {\rm erg\, s^{-1}\, Mpc^{-3}\, Hz^{-1}}\right) \nonumber\\
&\times&(1+z)^{4.68}\frac{{\rm exp}(-0.28z)}{{\rm exp}(1.77z)+26.3}.
\ea
The emissivity scales with $\nu^{-0.44}$ for ($\lambda>1300 {\rm \mathring{A}}$) and with 
$\nu^{-1.57}$ for ($\lambda<1300 {\rm \mathring{A}}$), as observed in the spectra of radio quiet sources 
\citep{2001AJ....122..549V,2002ApJ...565..773T}.
An upper limit to this emissivity can be obtained with the \cite{2015ApJ...813L...8M} model, derived from 
fits to several observational data sets. This upper limit is given by:
\be
{\rm log} \epsilon_{912}(z)=25.15e^{-0.0026z}-1.5e^{-1.3z}.
\ee
and scales with frequency as $\nu^{-0.61}$ \cite{2015MNRAS.449.4204L}.

The quasar luminosity density contributing to the UV background at the observed \lya frequency is 

\be
I^{\rm quasar}_{\nu_o}(z_{\rm Ly\alpha}) = 10^{-73.468}\int_{\nu_{\rm min}}^{\rm \nu_{\rm max}} d\nu 
\frac{\epsilon_{\nu}\, y(z)}{4\pi(1+z)^5},
\ee
where $\nu_{\rm max}=6.0\times \nu_o$. Note that the bulk of this emission originates at redshifts 
very close to $z_{\rm Ly\alpha}$ or at lower redshifts. Therefore, the result is not very sensitive to 
the choice of $\nu_{\rm max}$.

As observed in Figure \ref{fig:cont_z}, the intensity of continuum contamination in \lya intensity maps
is higher than the intensity of \lya emission from filaments. The former intensity is, 
however, quite uncertain, with model predictions differing by over one order of magnitude 
\citep{2011MNRAS.410.2556D}. Therefore, constraining this extragalactic background (continuum) light 
(EBL) is one of the objectives of the proposed experiment.
Since the SFR model we take as referrence is well within current constraints, we believe that the difference 
in the predicted EBL originates in the uncertainty in the quasars 
UV emissivity. This further supports the need to detect the effects of this UV continuum in the IGM, so 
that we can better constrain it. 

The intensity of contamination found is compatible with the predicted evolution of the EBL 
from \cite{2011MNRAS.410.2556D} if we assume that the escape fraction 
of photons with frequencies above the \lya line is very small. This would result in the 
contamination given by the black dashed dotted line in Figure \ref{fig:cont_z}. Our model can also easily 
fit the observational constraints from \cite{2000ApJ...542L..79G} and from \cite{2005ApJ...619L..11X} 
for different assumptions about the photon escape fraction.

However, the smoothness of this foreground across frequency, compared
to the \lya signal, should allow it to be fitted and removed from the observational maps in the 
same manner as the foregrounds for 21 cm line intensity maps are removed (e.g. \citealt{2006ApJ...650..529W}).
This fitted foreground can, in principle, be used to probe the EBL.

\section{Conclusions}
\label{sec:Conclusions}
Mapping gas filaments in the IGM is an essential step to complete our picture of the spatial distribution of 
large scale structures throughout the Universe.
These maps can provide much more information about the baryonic matter spatial distribution than
observing emission from galaxies, which are just discrete point sources in the point of view of large scale studies.

Cold IGM filaments can only be directly detected through emission lines from hydrogen or helium. Therefore, 
this study explores the prospects for detecting and mapping hydrogen \lya emission from IGM filaments at $z<3$. 

In this redshift range the \lya line is mostly observed in the UV band and, according to our conservative 
predictions, the \lya  signal from cold IGM filaments can be detected by an experiment with a sensitivity of 
${\rm 3.7\times10^{-9}\, erg\, s^{-1}\, cm^{-2}\, sr^{-1}}$ (${\rm \sim 37\, mag/arcs^2}$) in the UV band, 
which is in the reach of the next generation satellites. The \lya emission from these
filaments is powered by the UV background and, in most cases, will emit with a maximum intensity of about half 
the intensity of the UV background. It can therefore be used to probe and constrain this poorly constrained background.

Emission in the \lya line from filaments can be used to probe the baryonic content of the IGM, as well as 
the thermal and ionization state of these baryons.
Intensity mapping of IGM filaments with two emission lines, such as the HI \lya and the HI ${\rm 21\, cm}$ 
line, will allow for even better constraints to the astrophysical conditions in filaments. This is essential 
to properly model the gas flow between galaxies and the IGM.

In this study, we consider the continuum foregrounds that will contaminate \lya intensity maps and find 
that a mission with the experimental setup that we here propose has the required sensitivity to map 
the proposed signal and enough angular and frequency resolution to be able to remove the main line foregrounds 
and continuum foregrounds.

\section*{acknowledgements}
We thank the Netherlands Foundation for Scientific Research support through the VICI grant 639.043.006.

\bibliography{Lya}

\end{document}